\documentstyle[12pt,epsf,rotate,times]{article}

\newcommand\et{{\it et~al.}}
\newcommand\ie{{\it i.e.}}

\def\mathmode#1{\ifmmode {#1}
                  \else {$#1\mkern-5mu$} \fi}
\def\msun{\mathmode{\,M_\odot}}
\def\lsun{\mathmode{L_\odot}}
\def\mcore{\mathmode{M_{_{\rm c}}}}
\def\menv{\mathmode{M_{_{\rm env}}}}
\def\teff{\mathmode{T_{_{\rm eff}}}}
\def\lesssim{\mathrel{\hbox{\rlap{\hbox{\lower4pt\hbox{$\sim$}}}\hbox{$<$}}}}
\def\gtrsim{\mathrel{\hbox{\rlap{\hbox{\lower4pt\hbox{$\sim$}}}\hbox{$>$}}}}
\def\deriv #1#2{ {{{\rm d} \, {#1}} \over{{\rm d} \, {#2} }}}
\title{\bf Horizontal Branch Stellar Evolution }
\author{Ben Dorman$^{1,2}$\\
\vspace{1cm}\\
\normalsize $^1$Laboratory for Astronomy \& Solar Physics, NASA/GSFC,\\
\normalsize Greenbelt, Maryland 20771, USA \\
\normalsize $^2$NAS/NRC Resident Research Associate}
\date{\mbox{}}
\begin{document}
\maketitle
\pagestyle{empty}
%
%
\def\bull{\vrule height .9ex width .8ex depth -.1ex}
\makeatletter
\def\ps@plain{\let\@mkboth\gobbletwo
\def\@oddhead{}\def\@oddfoot{\hfil\tiny\bull\quad
``Stellar Evolution : What Should Be Done'';
32$^{\mbox{\rm st}}$ Li\`ege\ Int.\ Astroph.\ Coll., 1995\quad\bull}%
\def\@evenhead{}\let\@evenfoot\@oddfoot}
\makeatother
%
%
\def\beginrefer{\section*{References}%
\begin{quotation}\mbox{}\par}
\def\refer#1\par{{\setlength{\parindent}{-\leftmargin}\indent#1\par}}
\def\endrefer{\end{quotation}}
\def\ga{\mathmode{\nabla_{ad}}}
\def\gat{\mathmode{\nabla_{ad_{,T}}}}
\def\gap{\mathmode{\nabla_{ad_{,P}}}}
\def\gr{\mathmode{\nabla_{r}}}
\def\kapo{\mathmode{\kappa}}
\def\ydot{\mathmode{\dot Y} }
\def\cdot{\mathmode{\dot C} }
\def\tdot{ \mathmode{\dot T} }
\def\kappadot{\mathmode{\dot\kapo} }
\def\kff{\mathmode{\kappa_{\rm ff}}}
\def\kes{\mathmode{\kappa_{\rm ff}}}
\def\kapp{\mathmode{\kappa_{_{,P}}}}
\def\kapt{\mathmode{\kappa_{_{,T}}}}
\def\ldev #1#2{ {{\rm d} \, {\log #1}} \over{{\rm d} \, {\log #2} }}
\def\et{{\it et~al.}}
%
%
\vbox{
{\noindent\small{\bf Abstract:}
I review aspects of the evolution of horizontal branch (HB) stars.  The
defining characteristic of this, the core He-burning phase of evolution
for $M < M_{\rm flash} \sim 2.2\msun,$ is a spread in observed colour
amongst HB stars within the  (almost) chemically homogeneous,
isochronous populations found in Galactic globular clusters. The
inference is that the stars within a given cluster must have undergone
varying degrees of mass loss during the earlier, red giant (RGB)
phase.

I start by reviewing current topics in the study of HB stellar
evolution, including a brief review of the main determinants of the
structure of low-mass core helium burning stars and  of HB morphology.
I describe the main concerns addressed by earlier studies
of HB morphology, {\it viz.} the `first' and `second' parameter
effects on HB morphology, and attempt to summarize the current
state of affairs and how future investigations might improve
our findings in this area.
I also briefly discuss what appears to be the topic that has generated the
most theoretical attention over the last 5 years,
that of the hot end of HB sequences. The stars that have
undergone the most mass loss --- termed extreme HB stars (EHB) --- form
a group that is often separated observationally from the `blue' HB
stars seen in metal-poor globular clusters.  I then discuss the current
state of HB theory, reviewing the effect of new physics on the models,
and the special considerations (partial mixing) that arise in core
helium burning models.  Assuming only that the gas cannot, {\it on
average,} be strongly superadiabatic anywhere, one finds that the
convective core must be surrounded by a partially mixed
semiconvective region. This result holds when the radiation
pressure is small and the opacity has a significant Kramers-type
component.  I also review the controversy surrounding the behaviour of
the core in the late phases of evolution.

Finally, I give two further examples of questions that are still outstanding.
After noting that the mass loss mechanism is not at all
understood, I review detailed work by several different groups
concerning the age, abundance spread, and rotational properties
of RGB and HB stars in the  `second parameter' clusters M3 and M13.
I conclude by discussing  the HB luminosity-metallicity relation.
The discrepancy between the slope of this relation from different
lines of argument appears largely to have been resolved.
However, the observational data and theoretical sequences
(if $Y \gtrsim 0.23$) are still
in conflict as far as the HB luminosity zero-point is concerned.}
}

\newpage
%
%
\section{Introduction}

\subsection{Overview}

The helium-burning phase of stellar evolution has a particular
importance, because it produces two different `standard candles' of
pulsating stars at different mass ranges: Cepheids at intermediate
masses and RR Lyrae stars at low mass.  Core helium ignition produces
stars that spend significant time within a region just blueward of the
red-giant branch (RGB).  As the red edge of the instability strip is
related to the line on the HR diagram where stars no longer have
extensive convection zones, it follows that He-burning objects are the
longest-lived stars that occupy the pulsationally unstable region.

This review focuses on the low mass core helium burning stars that
appear as either `red clump' stars or horizontal-branch (HB) stars.
For a fixed composition, the HB is understood to be a sequence of
objects with fixed core mass \mcore\ and varying envelope mass \menv.
After attempts to fit observed HBs with sequences of constant mass, it
was realized  (Iben \& Rood 1970) that there was an intrinsic scatter
in the HB masses due to mass loss in prior stages, most probably on the
RGB rather than during the He-flash.  Thus the defining characteristic
of the HB is actually determined by a `non-canonical' (and
ill-understood) process that occurs prior to the phase itself. The
dispersion in mass along observed HBs is found to be of order a few
hundredths $\msun$ (Rood 1973; Lee, Demarque \& Zinn 1990), and it is
probable that most stars undergo some mass loss. The process underlying
mass loss is not identified, at least in cool, single, non-pulsating
giants, and neither  its dispersion or the shape of the mass
distribution (d'Cruz \et\ 1996) are understood. The dispersion is
likely to be the result of variations in individual properties amongst
stars, and it is found not to be the same in general amongst
different clusters; its cause  probably also contains a stochastic component.

\mcore\ is determined by the onset of the helium core flash; energetic
considerations, constraining the amount of energy required to lift the
central degeneracy, allow only weak variations among stars at fixed
abundance. Small differences may, however, arise among stars because
of varying rotation rates (Renzini 1977) or perhaps star-to-star
variations in initial CNO abundance (Briley \et\ 1994).
\mcore\ additionally varies little with age (for $t: M_{_{\rm RGB}} <
M_{_{\rm He-flash}}$), which is the property that in principle makes
the HB a good standard candle.

The models have almost fixed core luminosity, but widely varying
hydrogen shell luminosity (Iben \& Rood 1970; Dorman 1992). Their
hydrogen-rich envelope structure is `giant-like' at the red extreme,
where the hydrogen shell contributes most of the energy and forces a
large convective envelope.  For smaller envelope masses the H-shell
strength drops until the outer zones are radiative, and the sequence
crosses the instability strip.  The envelopes are `dwarf-like' at the
blue end, where the outer hydrogen envelope remains inert until the
He-shell-burning stage. The basic qualitative picture does not vary
much with metallicity, helium abundance or any of the other parameters
known to affect this phase of evolution.

\subsection{HB Morphology: the First and Second Parameters}

\begin{figure}[t]
\epsfysize 4.75truein
\hfil\epsffile{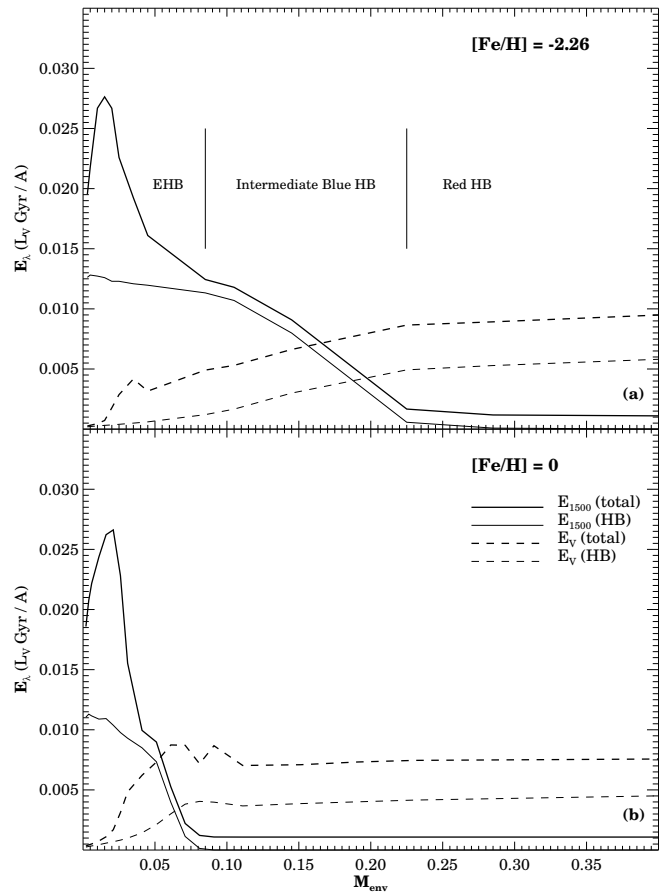}\hfil
\caption{\label{fig:energy} Lifetime energy radiated at 1500 \AA\ (solid)
and in the $V$ band (dashed) plotted as a function of envelope mass \menv.
The solid thick lines include the radiation up to the end of the post-AGB
phase. Thinner lines give the same quantity for the HB phase only. The
height of far-UV curve is a measure of the mean temperature of the stars
during evolution. For the red HB stars, the UV-flux is all produced
in the post-AGB phase.}
\end{figure}

\begin{figure}[t]
\epsfysize=4.75truein
\hfil\epsffile{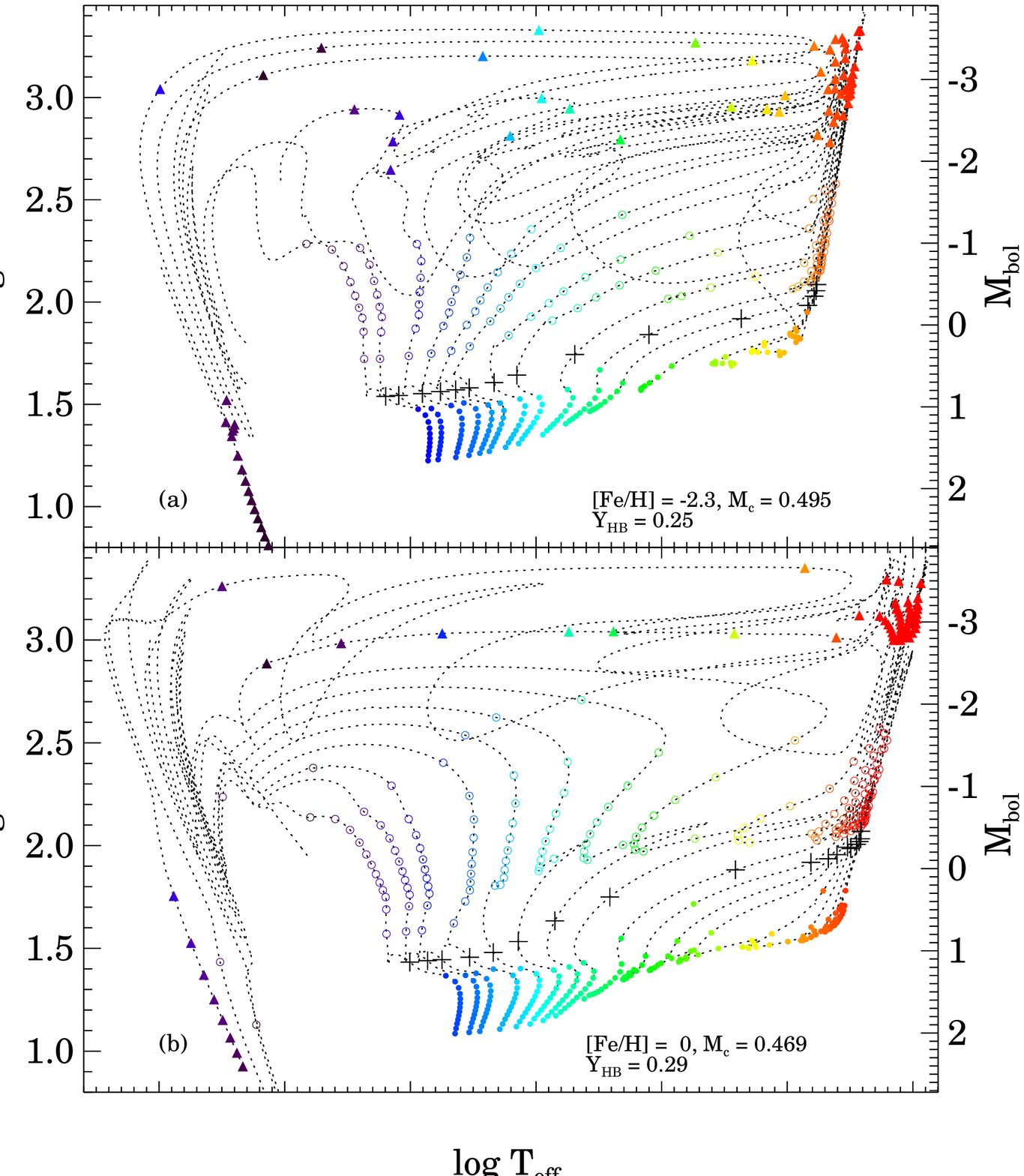}\hfil
\caption{\label{fig:tracks} HB evolutionary sequences for [Fe/H] $= -2.26$
and 0, from Dorman \et\ 1993. The smallest envelope mass in each set
is 0.003 \msun. Filled circles are placed every 10 Myr in the HB phase; open
circles every 2 Myr in the post-HB phases; filled triangles are placed
every 0.1 Myr from the point where $\log L/\lsun$ reaches 3. The `wild'
oscillations are due to helium \& hydrogen shell flashes subsequent to the AGB
phase.}
\end{figure}

Fig.~\ref{fig:energy} illustrates the $\menv-\teff$ relationship
for HB stars in a somewhat unconventional fashion. I plot the lifetime
radiated far-UV (1500 \AA)  and $V-$band energy against the mass
\menv\ of the hydrogen-rich envelope, for models with $\rm [Fe/H] = 0,
-2.26.$ The corresponding evolutionary tracks (Fig.~\ref{fig:tracks})
are from Dorman, Rood, \& O'Connell (1993). The ordinate in
Fig.~\ref{fig:energy}
turns out to be  an important quantity for population synthesis studies: see
Dorman,
O'Connell, \& Rood (1995) for a discussion. Since only stars with $T \ge
7500~{\rm K}$ radiate significantly in the far-UV,  the $E_{1500}$
curves measure the luminosity- and time- averaged mean temperature.
The $V-$band curves are flat for the range of masses that corresponds to
the red clump.  Both high and low metallicity models with small
envelopes are extreme HB (EHB) stars, defined by the criterion that
they are hot during post-HB phases, i.e. that they never reach the asymptotic
giant branch (AGB).
Evidence from the Galactic field as well as the globular clusters
suggests that these might be produced by mass loss processes distinct
from those responsible for the cooler stellar distribution.
In the metal-poor case there is a range of `intermediate' blue stars
that correspond to the globular cluster blue HB stars.  For metal rich
populations, models with similar \menv\ to the blue HB stars are red,
and the temperature range including the instability strip and the blue
HB is unlikely to be populated.

This is, of course, simply the `first parameter' effect; all HB
morphologies are possible according to Fig.~\ref{fig:tracks}, but some are much
more likely at low
metallicity (Fig.~\ref{fig:energy}) because the mass range that allows them is
larger and the
mass loss required to produce them is less extreme.  This figure also
encapsulates the `intrinsic' effect of stellar interior physics on HB
morphology. The so-called `second parameter'  effects --- variations in
HB morphology at  (apparently) fixed abundance include all the `stellar
populations' considerations, which may be thought of as all the
phenomena that are larger than the individual stars. For example, if
individual characteristics such as rotation affected the HB morphology,
the question to be answered would be why the stars in one cluster have
such different mean rotational properties from another.  In this sense,
the rather well-researched hypothesis (Zinn 1980; Lee, Demarque \& Zinn
1994; Fusi Pecci \et\ 1995) that age is the `global' second parameter
determining HB morphology is highly satisfactory, as it would explain
in a most natural manner why there are differences among apparently
similar clusters. This idea is reinforced by the prevalence of the
problem in clusters outside the Solar circle.  A recent theory of
globular cluster formation (McLaughlin \& Pudritz 1995) suggests that
formation timescale for globular clusters varies with galactocentric
radius, consistent with this interpretation of HB morphology variations.

However, any study of the HB stellar distribution necessitates
simplifying assumptions that are  difficult to justify a priori, since
we do not have a viable theory for the mass loss process. Fig.~7 of Lee
\et\ (1994) demonstrates a convincing case for age as the second
parameter; for globular clusters with Galactocentric radius $8 < R <
40\:  {\rm kpc},$ at fixed HB type [as parametrized by the Lee index
$(B-R)/(B + V + R),$ where $B, V, R$ are the numbers respectively of
blue, variable and red HB stars] the outer halo objects are more
metal-poor, or alternatively they have bluer HB's at fixed
metallicity. Assuming that the mass loss is invariant with abundance,
at least for $\rm [Fe/H] > -2$ (below which the HB morphology is not
monotonic), synthetic HB models of different ages fit the cluster data
implying a trend toward bluer HB morphology at greater age.  This argument
was one of the motivators behind the widely adopted Searle \& Zinn
(1978) picture of galaxy formation. It should be pointed out that
another assumption is implicit in the interpretation of the HB data.
Assuming that mass loss is fixed in either time and/or metallicity also
implies that the intrinsic stellar mass distribution is the same in different
clusters.  Thus, comparing clusters at the same metallicity for age
differences requires verifying that the mass distributions are drawn
from the same distribution. This is not the case, for example, between
NGC 6752 and the `young' globular cluster Ruprecht 106.  Rigorous tests
of mass distributions for different clusters have not, to the author's
knowledge, often been considered to date (but see Dixon \et\ 1996 for
the case of M79). In addition, the Lee index is insensitive to HB
variations if the stars all lie blueward of the instability strip. This
is unfortunate since we are interested in placing bounds on the ages of
the oldest clusters.  More studies and tests of the mass distribution
are a crucial addition to this field, requiring large samples
of data and goodness-of-fit tests.

\subsection{The Horizontal Branch and UV-Bright Stars}

\begin{figure}[t]
\epsfxsize=3.0truein
\hfil\rotate[l]{\epsffile{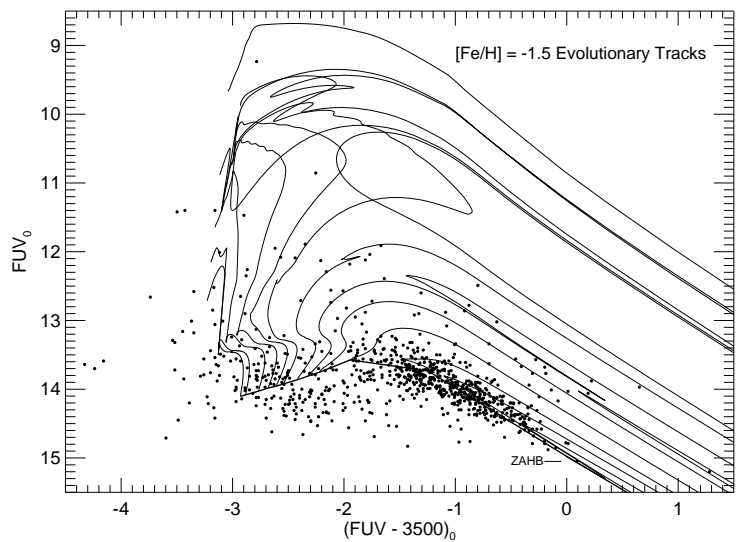}}\hfil
\caption{\label{fig:wcen} A far-UV--near-UV colour-magnitude
diagram of the globular cluster $\omega$ Centauri (from
Whitney \et\ 1994). Superposed are
tracks from Dorman \et\ 1993. The redder stars
correspond to blue HB stars seen in the HB blue tail.
The bluest stars
appear to form a separate clump.}
\end{figure}

In the last 5  years a fair proportion of the theoretical work on HB
stars has focussed on the blue extreme members of this phase (Brocato
\et\ 1990; Castellani \& Tornamb\`e 1991;  Horch, Demarque \&
Pinsonneault 1992; Dorman, \et\ 1993; Bressan, Chiosi, \& Fagotto
1994). The reason for this is that extragalactic stellar populations
that have no sign of recent star formation or nuclear activity are
invariably detected in the far-ultraviolet (Code \& Welch 1979;
Burstein \et\ 1988).  HB stars have gained a new significance for
stellar population studies as they are potentially the largest
contributions to the integrated light of a galaxy at vacuum ultraviolet
wavelengths.  They are both reasonably bright, and very long lived
compared to other candidates from old populations {\sl viz.}
Post-AGB stars, of which the Planetary Nebula
Nuclei form a subset.  Fig.~\ref{fig:energy} makes it clear
that both hot HB and EHB/AGB-Manqu\'e stars produce UV flux
well in excess of the  Post-AGB stars. The figure emphasizes
that {\it all}
of the following categories of stars are UV-bright: (a) blue HB stars,
as found in metal poor globular clusters (Watson \et\ 1994); (b) EHB
stars, as found in the Galactic disk as subdwarf B (sdB) stars (e.g.
Saffer \& Liebert 1995; Heber 1992); (c) AGB-Manqu\'e stars, which
probably correspond to the subdwarf O stars (e.g. Dorman \et\ 1995 and
references therein).  These latter are the post-HB (He shell-burning)
descendants of the EHB stars. The largest sample of EHB stars observed
in a globular cluster is that found in $\omega$~Centauri
(Figure~\ref{fig:wcen}), which is thus not only the largest globular
cluster but also turns out to have the bluest HB.

Figure~\ref{fig:wcen} shows
the color-magnitude diagram of Whitney \et\ (1994), from the Ultraviolet
Imaging Telescope (UIT) that flew on the Space Shuttle in December 1990.
The space observation has been combined with ground-based Stromgren
$u$ photometry covering the centre of the cluster, which is completely
resolved in both. The observations are overlaid with theoretical tracks from
Dorman \et\ (1993). The redder stars in the diagram ($\rm FUV-3500 > -1.5$)
correspond to the HB tail as seen in clusters
such as M13, while the bluest objects appear to form a separate clump.
An unresolved problem is their relative faintness compared to the
theoretical sequences; in contrast the cooler stars fit the
theory within the photometric errors $\rm (\, \sim 0.1\: mag).$
This problem may be resolved by HST
Cycle 5 observations  currently in reduction (Whitney \et\ 1996, in
preparation). The apparent separateness of the bluer stars may imply
that a different mass loss mechanism is involved in their
creation. The large numbers imply a single-star origin rather
than one due to binary interactions. The same separation
is observed between the Galactic field hot subdwarfs and the field blue HB
(Greenstein \& Sargent 1974; Heber 1992), and also in the globular cluster
NGC~6752
(Buonanno \et\ 1986).

\subsection{Outline of this Paper}

The question we wish to address here is `What theoretical work needs to
be done in order to improve our understanding of the phase and/or
agreement with observations?' To answer this question,
 I start by reviewing the effect of new physics
on HB models; for reasons that are clear, the effect is limited,
except for the study of RR Lyrae masses. I then review
the evolution of the convective core and the controversial question
of the late time evolution and the so-called `breathing pulses'
found in models close to helium exhaustion. These are a source of
some uncertainty in population synthesis applications. I adopt the
view here that `canonical semiconvection' provides a description
of the core evolution that is consistent with observation, and in turn
implies that, while the core evolution may not be completely smooth,
the larger manifestations of the breathing phenomenon found in evolution
codes are artificial.

The most overwhelming uncertainty, however, is that of mass loss in red
giant stars, because it affects not just the fine details  but the
gross properties of the stars. Of course, such phenomena are not
strictly speaking within the scope of this paper. However we will
summarize some recent work on the possible relation between individual
stellar properties (rotation, surface composition) on horizontal branch
morphology in section~\ref{sec:RGB}.  Another source of uncertainty
that needs to be resolved is the relation between HB luminosity and
metallicity (see also the review by Chaboyer, this volume). This
represents a potential serious problem either in our understanding of
the HB phase or in the evolution of Galactic chemistry.  We will
address this in section~\ref{sec:mbolfeh}. The reader is also referred
to the excellent review of theory and its comparison with observation
by Renzini \& Fusi Pecci (1988).

\section{The Current State of the Theory}

\subsection{New Physics \& Null Results}

The new generation of opacity calculations by the OPAL and OP projects
(Rogers \& Iglesias  1992; Seaton \et\ 1994), which were spurred
by the Cepheid mass discrepancy (Simon 1982), have made significant
breakthroughs in the reconciliation of pulsation and evolution theory
(see the conference volume by Nemec \& Matthews 1993).  The Cepheid
masses are now in agreement, and the RR Lyrae mass discrepancy (Cox,
Hodson \& Clancy 1983) has been rendered insignificant.  Yet their
impact  on the gross properties  of the HB stars themselves is small
(Dorman 1993; Yi, Lee, \& Demarque 1993).  Why is this? Major revisions
in the opacities have significantly increased the opacity due to iron
at temperatures of a few hundred thousand degrees. However, the opacity
deep in the stellar interior has not been greatly revised. The
luminosity of HB stars is determined by the \mcore and by the envelope
opacity, which both act to regulate the hydrogen burning shell (Dorman
1992). It is currently thought that the opacity of elements that are
either fully ionized or ionized to the hydrogenic level are fairly well
understood, so that no major revisions are expected.  Since the opacity
function within a star is typically dominated by minority elements, one
may well ask whether some unaccounted for opacity source may yet shake
the current accord between the two major opacity `providers.' The
answer may well be negative simply because suitable candidate elements
are simply too rare (see Iglesias \et\ 1995), but this is one remaining avenue
for revisions in
the physics that might cause changes in the evolution.  However, such
changes may also seriously affect age estimates made using theoretical
isochrones.

Unlike the case with the opacity
the equation of state (EOS) is dominated by the majority ions, hydrogen and
helium. This is because
the effects of different elements add harmonically (Fontaine \et\ 1977)
through the `additive volume' law, strictly correct for non-interacting
gas species by application of the First Law of Thermodynamics.
Thus, simple models of the gas give good results until carbon
and oxygen dominate as constituents, later in the HB phase.
The main
additional consideration is the consistent inclusion of non-ideal effects,
chiefly Coulomb forces and the phenomenon loosely referred to as
`pressure ionization,' by which elements stay ionized at high pressures
contrary to the prediction of the ideal gas Saha formula. The problems
involving the latter are not specific to helium burning. However simple
calculations show that, while the gas in the He-burning cores is
close to ideal (the Coulomb parameter $\Gamma < 0.1$) at the ZAHB,
the carbon and oxygen abundances increase with burning so that $\Gamma
\sim  2$ at later stages of evolution. Unfortunately we do not yet have
an EOS that is appropriate for a partially relativistically degenerate
gas with relatively strong Coulomb interactions. However the fact that
the gas is close to ideal (and non-degenerate, for that matter) close
to the ZAHB implies that EOS effects will not change the luminosity
significantly. The track morphology may however be affected, modifying
the conclusions to be drawn from detailed studies of HB morphology. But
it seems that such modifications will change nothing except
the quantitative details of stellar mass distributions.

For the reaction rates the greatest source of uncertainty has been the
$\rm ^{12}C(\alpha,\gamma)^{16}O$ rate. The rates given by Caughlan
\et\ (1985) and Caughlan \& Fowler (1988) differed by a factor of 3.
The earlier, higher rate gives terminal C:O ratio 1:4,  whereas the
later rate implies the same ratio is nearly 1:1. The difference may be
important for the study of white dwarf cooling times, but the effect on
HB morphology is once again slight, involving a change in the precise
temperature width of blueward loops (Dorman, Lee, \& VandenBerg 1991).
Weaver \& Woosley (1993) determined from a study of nucleosynthesis
yields in supernov\ae\ that the `true' rate of this reaction was
($1.7\pm 0.5$) times the Caughlan \& Fowler (1988) rate. Recent
experimental work (Arnould, these proceedings) appears to corroborate
this finding.

To summarize, one can scarcely do better today than to quote Iben and
Renzini (1984, p.357): {\em ``We do not infer that standard stellar
models are therefore correct. We do note, however, that there are many
quantitative matches between standard model characteristics and cluster
characteristics ... and caution that changes in the stellar physics
that are invoked to produce non-standard models at some desired age may
possibly destroy the good matches with other cluster characteristics
that now exist.''}

\subsection{Evolution of the Convective Core}

\subsubsection{A Criterion for `Semiconvection' or `Partial Mixing'}

\noindent As is well-known, helium burning stars develop convective
cores  very close to the onset of central helium burning. The opacity
increases in the convective core because of the carbon enrichment
produced by the nuclear reactions. After the nuclear-processed material
has been cycled to the outer boundary of the core, it will have cooled
enough to be unstable to convection.  The convective region  must
therefore penetrate into the overlying radiative layers (see Renzini
1977) and the core will grow so that it retains convective stability at
its outer boundary, additionally taking in unprocessed He.  A time is
reached, however, when the core expands to a point such that engulfing
fresh helium does not stabilize its outer layers.  We determine
stability by the Schwarzschild (1906) criterion (one may also suppose
that the core overshoots the boundary determined by this criterion, to
a point where the gases are decelerated to zero, but this should not
qualitatively affect the argument).  At this time, instead of decreasing
beyond the edge of the core, {\em for a homogenous mixture} the gas is
instead ``more'' convectively unstable owing to a local minimum in the
ratio $\gr/\ga$ at the boundary. Here, I summarize  the
criterion of Dorman \& Rood (1993) for the size the convective core
must attain before the critical point is reached.

Define

\begin{equation} \label{eqn:r}
R = {\gr \over \ga }\propto { { \kapo L P}\over {MT^4\ga}},
\end{equation}

\noindent where the variables $\kappa, L, P, T, M, \ga, \gr$ take their usual
meanings and are functions of radius.  We seek the point where the
derivative $D = {{d\, \log R}/{d\,  \log r}} = 0.$ This point will have
significance only if, in addition, the stability criterion  is
satisfied where the derivative changes sign, \i.e.\ if $D = 0$ where $R
= 1$ at some point during the evolution. Denote this critical
point by $P.$ The derivation is similar to
that of Naur \& Osterbrock (1953), except that no explicit analytical
form is substituted for the solution.  Instead, the resulting
expression locating $P$ can easily be used in a stellar evolution code,
as one needs to compute all of the necessary quantities for other
purposes.

Recall the definitions of the Homology Invariants $U, V, W$
 (Schwarzschild 1958):

\begin{equation}
 U ={\ldev {M} {\it r} }, \hskip 10pt V = -{\ldev P {\it r} },
\hskip 10pt W = {{\ldev {L} {\it r}} \sim
\varepsilon_{therm}+\varepsilon_{nuc}-\varepsilon_\nu }
\end{equation}

\noindent Differentiation of (\ref{eqn:r}) yields

\begin{equation}
D = U - W +  V{\bigr[ 1 + \kapp -  \gap + \ga (\kapt - 4 - \gat)\bigl]}
\end{equation}

\begin{figure}[t]
\epsfysize=3.0truein
\hfil\epsffile{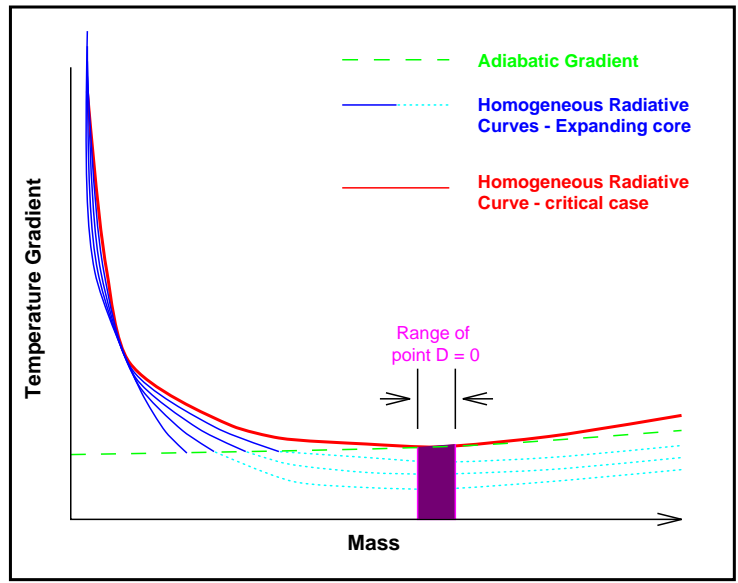}\hfil
\caption{\label{fig:dcore} The change in the radiative gradient with
increasing abundance for a homogeneous composition. The gradient ratio
drops at the very center but rises outward, increasing the size of the
core until it attains the point $P: D=0.$ The  point $P$ is a function
of $U$ and $V$ and quantities that vary slowly, so that it remains roughly
fixed in mass.}
\end{figure}

\noindent  Consider the terms in this equation.  Away from the centre
$\varepsilon_{nuc} \approx 0$, so $W$ is small. The derivatives of
\ga\ and \kapo\ with respect to pressure and temperature can be
qualitatively summarized as follows: \gap\ and \gat\ are small away
from ionization zones and where the radiation pressure is small, and
\kapp\ and \kapt\ are slowly varying with time. In fact since we can
write the opacity as the sum of free-free (Kramers-like) and electron
scattering terms, with the former contributing about 20\% of the total
opacity, we can show that the opacity derivatives also change slowly.
It should be emphasized that only a relatively small change in the
opacity is sufficient to drive the core boundary outward.  Thus $P$
tends to occur at a slowly varying value of $U/V$ throughout evolution.
In addition $U$ and $V,$ being homology invariant, do not change with
composition at a fixed mass co-ordinate provided the core remains
convective. Since the core initially grows monotonically in mass with
increasing carbon abundance, the  point $P$ will encountered at
some time during the evolution. The situation is illustrated in
Fig.~\ref{fig:dcore}, which shows schematically the radiative and
adiabatic gradients as a function of mass. The key point is that for any core
composition
there is a point where $D=0,$ because it is approximately fixed in $(U, V)$
coordinates. It becomes a critical point  only when the core has
grown sufficiently to encompass it.

\subsubsection{Partial Mixing Zones}

What will happen after the critical point is reached? If the gas must
obey the stability criterion, then the convective core cannot expand to
produce a larger fully mixed core, nor can it stay the same size.  The
operation of any of the processes proposed to induce convective
overshooting will be much faster than the nuclear timescale. Thus of
most interest to modelling the evolution is what will be the {\it mean}
long-term behaviour of the carbon (and later oxygen) enriched gas that
moves outward. The minimum degree of mixing possible under the
assumption that the gas is, on average, subadiabatic or neutrally
stable, is that material reaching the convective core boundary will
continue to move until it has mixed with its surroundings sufficiently
in order to be stable. This is the physical reasoning behind the
`canonical semiconvection prescription,' in which it is assumed that
the gas achieves convective neutrality beyond $P.$ Now the
undisturbed, almost pure helium mixture beyond the convective core
tends to be strongly radiative, and mixing fully with the more carbon
rich core would make it convectively unstable. It follows that if the
mixture above the core must have $\gr \approx \ga,$ it will
reach equilibrium at an intermediate composition. The actual  composition
at neutrality is dependent on the local conditions, implying
a composition gradient. This argument holds as long as $\ga$ is not
strongly affected by changes in the radiation pressure (so that the
core remains homologous with time) and the opacity has a
significant component which is Kramers-type (\ie\ the opacity is not
overwhelmed by
scattering processes). My calculations indicate that all stars with $M
\lesssim 8 \msun$ must contain a partial mixing zone; whatever
processes do occur outside the convective core must be mediated by such
a partially mixed region.

The attractiveness of this picture is that we do not need to know what
process causes the overshooting to determine the composition
profile. The constraint that the mixture should
be stable on timescales longer than the convective timescale implies a
`thermodynamic' constraint on the structure of the core, in the sense
that whatever the details of the processes, the structure of the core
must obey certain rules.  However, the details may well be mitigated by
the effects of overshooting beyond the boundary. Also, depending on the
efficiency of mixing processes, the evolution may or may not be smooth
on short timescales. Our evolution codes are not sufficiently detailed
to address such questions, especially as spherically symmetric 1-D
calculations can only model global perturbations. Since `canonical
semiconvection' is also consistent with the ratio $R2$ of HB stars to
AGB stars (Dorman \& Rood 1993; Renzini \& Fusi Pecci 1988), it is
tempting to conclude that this mechanism for regulating the flow of
helium into the core provides a reasonably accurate `zeroth order'
approximation to the true evolution of He burning cores. The theory
requires no free parameters except that convective overshooting is
efficient enough to allow the required mixing---a very weak condition.

\subsubsection{Late He Burning Evolution \& the Breathing Pulse `Phenomenon'}

Relatively late in the core He-evolution ($Y_{_c} < 0.08$, increasing
to $\sim 0.2$ for $M \sim 3 \msun),$ converging to the chemical profile
which has neutral convective stability apparently becomes more
difficult.  Evolution codes tend to produce core `breathing pulses'
(BPs), in which the core expands suddenly in mass and radius, the
central temperature drops, the central helium abundance rises, and as a
result the direction of evolution is temporarily reversed. However,
inspection of the terms in \gr\ shows that none of them have a
tendency to cause an increase in the core size at late times. In
particular, the opacity that drives the core boundary outward earlier
in the evolution must start to decline before core exhaustion is
reached.  Assuming that the opacity is part free-free and part electron
scattering, its  time dependence can be written

\begin{equation}
{\biggl({{\partial \log \kappa} \over {\partial \, t}}\biggr)_M } =
\biggl({{-3\ydot - \cdot }\over{4 - 3Y - C} }\biggr) + \bigl(\kapp/\ga +
\kapt\bigr){\tdot \over T}. \label{eqn:ydot}
\end{equation}

\noindent for helium abundance $Y$ and carbon abundance $C,$ and where each of
the time derivatives of composition are
composed of two terms, e.g. for helium

\begin{equation}
 {\deriv {Y} {t}} = \biggl(\deriv {Y} {t}\biggr)_{\rm mix} + \biggl(\deriv {Y}
{t}\biggr)_{\rm nuc},
\label{eqn:ymix}
\end{equation}

\noindent signifying the effect of mixing and of nuclear reactions. It
is clear that the two terms in equation~\ref{eqn:ydot} are of different
sign in general.  The coefficient of $\tdot$ is negative ($|\kapt| \ge
2.5|\kapp|),$ whilst the other term is
positive, as the opacity increases with decreasing $Y.$  Again,
the terms in equation~\ref{eqn:ymix} are also opposite in sign: nuclear
reactions will always decrease the He abundance, while mixing will
tend to increase it.

As $Y_{_c} \rightarrow 0$, clearly the composition changes must tend to
zero and the opacity must start to decrease. Thus at some point before
exhaustion, the opacity derivative at a fixed point in the model tends
to zero. Dorman \& Rood (1993) argue that it is this circumstance that
makes the end of core helium burning difficult to model.  They suggest
that the attempt by a partial mixing routine that depends on the
derivatives of opacity to produce a neutrally stable mixture is
problematical, because as the code attempts to find a solution with
${d\, \log \kapo /d\,\log t} = 0, $ it may find one with $\tdot$
negative and positive $\ydot$ -- which gives rise to a classic instance
of the BP phenomenon. It is not, however, clear that
this is correct, since various different iteration schemes all produce the
pulses.  Note however that the BPs need to be induced by localized
mixing events rather than by a global phenomenon, since the mixing
itself must be driven by the sum of such local events.  The process is
impossible to model with a 1-dimensional code; unfortunately, more
realistic simulations of convective cores still seem a long way off.
The case of BPs is  quite unlike that of the helium shell burning
thermal pulse phenomenon on the AGB, where the evolution of
global quantities brings about a thermal runaway (Schwarzschild \& H\"arm
1965).
In addition, the BPs have the following numerical characteristics: they are
not repeatable from model to model, and they appear discontinuously: a
`normal' model is followed by one with a large change in composition. Both of
these are untrue of the thermal pulses, which are highly repeatable and
manifest themselves by a steadily accelerating absorption of thermal
energy within the star as the helium shell grows exponentially in power
output.

There is, however, indirect evidence of irregularities in the evolution. There
are
well-documented cases of period decreases in RR Lyr\ae\ stars (see
Silbermann \& Smith 1995 for a recent detailed examination); several
have been observed using data collected over the last hundred years.
However, the connection with the breathing pulses is not firmly
established. The period change timescale is much shorter than the
timesteps used in any evolutionary model; the period `noise' predicted
from BPs has never been studied in sufficient detail to establish the
connection. Finally, the larger pulses that have been modelled in
evolution codes would produce changes in period that are far larger
than anything that has been observed.  Possibly, convection in the core
is not totally smooth in its operation, and mixing occurs spasmodically
(see Sweigart 1990). In the absence of other phenomena in the variable
star envelope that may affect the radius, core instabilities may
explain the period changes. However, the question remains open, and is
impossible to decide without detailed models of the core mixing
mechanism.

\section{RGB Mass Loss: Whence the Dispersion? \label{sec:RGB}}

As has been stressed earlier, the most fundamental missing piece to our
understanding of the HB phase is the nature and physics behind mass
loss (Fusi Pecci \& Renzini 1976). It is clear from the observational record
that both the mean mass
loss and its dispersion varies among clusters. Mass loss in red giants
is observed as H$\alpha$ emission wings (Cacciari \& Freeman 1983), but
such observations do not at present provide mass loss rates. As Renzini
\& Fusi Pecci (1988) have noted, the best constraints on RGB mass loss
come from the HB mass distribution itself, which is not entirely
helpful here.  The underlying mechanisms for mass loss must include
detailed study of such phenomena as molecular line driven winds and the
effects of grain opacity (Holzer \& MacGregor 1985; MacGregor \&
Stencel 1992). Briefly, in order to be effective, such sources for
coupling the stellar atmosphere to the radiation field must be
operative relatively close to the photosphere in order to produce
appreciable mass loss.

In this  section, I discuss another aspect of the problem: the
connection between variations in the properties of individual stars in
clusters with widely different HB morphologies. The question is whether
we can demonstrate some variation among properties of red giants that
corresponds to a difference on the HB. Fortunately, there is now one
set of clusters, M3 and M13, that can serve as a case study, although
(as might be expected) the results are not yet conclusive. The globular
cluster M13 has the `blue tail' HB morphology, with very few stars in
the instability strip and with the HB reaching to the main sequence
turnoff magnitude in the $V-$band. In contrast, M3 has many variables
and its HB extends from the red clump to $B-V \sim 0,$ well blueward of
the instability strip but only to $\teff \approx 10000 \: {\rm K}.$ By their
technique of shifting the cluster fiducial main sequence/subgiant/giant
sequences to determine differential ages, VandenBerg, Bolte, \& Stetson
(1990) concluded that the clusters do not differ in age more than 2
Gyr, a conclusion supported by the recent work of Catelan \& de Freitas
Pacheco (1995).

A complete sample of the brightest red giants in these clusters have
been studied spectroscopically by Kraft \et\ (1993).  Detailed
abundance analysis using a number of iron lines implies that their
metallicities are indistinguishable: $\rm [Fe/H] = -1.49$ in both
clusters (see their Fig. 2). However, analysis of their oxygen
abundances gave the surprising result that many of the M13 giants are
very oxygen poor, with $\rm [O/Fe]$ as low as $-0.8;$ similarly low
abundances were not found in M3. The inference is that the super-O-poor
stars dredge up layers that have been strongly affected by CNO
processing. In order to do this however, some mixing mechanism must act
far below the lower boundary of the convective envelope (e.g. Sweigart
and Mengel 1979; see also the recent work of Smith \& Tout 1992). The
preponderance of the anomalous stars close to the RGB tip also suggests
that the final mixing occurs very soon ($< 2$ Myr) prior to the
He-flash.

An obvious `individual property' responsible for variations in the
mixing and therefore the abundances is rotation. Peterson, Rood, \&
Crocker (1996) have obtained both oxygen abundances and rotational
velocities for 29 stars in M13 and 22 stars in M3. The object of their
study was to determine whether there was a relation between the oxygen
abundance or the rotational velocities on HB location. In M13, in the
temperature range where they had good sensitivity to O, they found no
super-O-poor stars. This suggests that they have
undergone large amounts of mass loss and are confined to the extreme
blue tail of the HB.  However, in both clusters the oxygen abundances
inferred show no strong trend with temperature although both samples
are confined to a narrow strip of the HB.  As far as the rotation was
concerned they found no obvious dependence of $v \sin i$ on HB colour.
The fastest rotators, found in M13, had $v \sin i \sim 40\, {\rm
kms^{-1}}, $ and the distribution of $v \sin i$ is bimodal. The
rotation velocities of M3 stars were found to be $\sim 10\, {\rm
kms^{-1}},$ with no  trend with temperature. The overall difference
between the global rotation/O abundance patterns between M13 and M3
leaves open the possibility that rotation may play some role in HB
morphology. With the current data there is no evidence that rotation
determines HB location on a star-to-star basis in a given cluster.
However, given that any correlation is likely to be quite noisy, a
definitive result must await observations of rotation spanning the full
range of the HB.

\section{ M$_{\rm V}$ vs [Fe/H]: The Zero Point Uncertainty
\label{sec:mbolfeh}}

In this final section, I discuss one of the `classical' problems with
of the HB, {\it viz.} the relationship between its luminosity and
metallicity. The properties of the HB play a key role in the
determination of the ages of globular clusters. Not only does the
`horizontalness' and theoretical invariance of the HB with age make the
HB sequence an important standard candle, but the fact that it may
contain the RR Lyr\ae\ instability strip reinforces this status by
allowing (in principle at least) an independent check on HB parameters.
A separate but related question is the so-called Sandage Period Shift
Effect, which can is an apparent inconsistency between
the observed pulsational properties of RR Lyr\ae\ stars and those
predicted from static theoretical models.
The literature on both of these questions is extensive, and I here try
to summarize some of the more recent work. Useful references are the
review by Sandage (1986), and the detailed discussions by Buonanno,
Corsi, \& Fusi Pecci (1989); Lee, Demarque, \& Zinn (1990), Sandage
(1990); Sandage \& Cacciari (1990) and a rigorous reanalysis by Fernley
(1993).

The data for mean HB magnitudes for different clusters are fit to a
linear relation of the form

\begin{equation} \label{eqn:mbz}
\rm \langle M_V \rangle = \alpha [Fe/H] + \beta.
\end{equation}

\noindent  In the last 3-4 years a consensus has emerged that the slope of the
observed relationship $\alpha$ is small, in the range 0.15--0.20. This
is in excellent agreement with the slope of theoretical sequences (Lee,
Demarque, \& Zinn 1990; Dorman 1993). There is some controversy about
the value of the zero point $\beta.$ This zero point affects age
determinations of clusters through the use  of $\Delta V^{^{\rm
TO}}_{_{\rm HB}},$ the magnitude difference between the main sequence
turnoff and the HB, as an age indicator (Chaboyer, this volume;
Chaboyer, Demarque, \& Sarajedini 1996).

Baade-Wesselink (BW) analyses also  give a small slope in equation
(\ref{eqn:mbz}), $\alpha = 0.15-0.21.$ (Carney, Storm, \& Jones 1992;
Clementini \et\ 1992; Fernley 1994). This is in contrast to typical
results of the main sequence fitting technique, which are derived using
the so-called `theoreticians route to distances' (Sandage 1986;
Buonanno \et\ 1989).  This method involves fitting theoretical
isochrones (as opposed to empirical main sequences) to cluster data in
order to derive distances.  The slope derived by Sandage and Cacciari
(1990) was $0.39 \pm 0.14.$ Sandage's analysis of the period-shift data
also implies a steep slope (Buonanno \et\ 1989). That is, he finds
(Sandage 1982)

\begin{equation} \label{eqn:spse}
\Delta \log P = -0.116 {\rm [Fe/H],}
\end{equation}

\noindent which, when taken with the van Albada \& Baker (1971)
pulsation equation, ($\log P \propto 0.84 \log L$ at fixed \teff),
yields $\alpha \sim 0.35.$ This is the connection between these two
classical problems.

Carney \et\ (1992) reanalyzed the results of main
sequence fitting. By demonstrating the existence of a metallicity
dependent error in the theoretical $\teff-(B-V),$ they were able to
reconcile the data to a slope $\alpha = 0.16.$ Fernley's (1993) analysis of
a carefully selected sample of RR Lyr\ae\ stars and temperature
determinations using optical/IR colours yield a much lower period shift
with metallicity (a coefficient of 0.073 in equation \ref{eqn:spse})
which reduces the slope to about 0.2.

But the BW analyses also consistently find a zero point
$\beta = 1.01 - 1.05.$ In  contrast, theoretical ZAHBs (Lee, Demarque,
\& Zinn 1990) give values $\beta = 0.82, 1.00$  for $Y_{\rm ZAMS} = 0.23,
0.20$ respectively (recall that $Y_{\rm HB} \approx Y_{\rm ZAMS} + 0.02$ due to
the first dredge-up). The profound implication of the controversy is that
the fainter zero point implies, using current theoretical models, a
lower value for $Y_{\rm ZAMS}$  than is
consistent with estimates from primordial nucleosynthesis and
worsening the disagreement between the implied expansion age of the
universe and that of the globular cluster system. For the reasons
explained in \S 2, there is relatively little room for manoeuvre
in the theoretical relations; the (correctly observed and interpreted)
magnitudes should allow a measure of the helium abundance of the
globular cluster system and should be consistent with other techniques,
in particular the $R-$method (see Buzzoni \et\ 1983). My bias lies
toward the suggestion that there is some systematic error in the methods
that obtain the fainter distance scale, as also suggested by
 Castellani \& de Santis 1994;
the reader is referred to the review by Chaboyer for a different viewpoint.

Consistent with the `faint' RR Lyr\ae\ distance scale are statistical
parallax analyses (Barnes \& Hawley 1986). They found

\begin{equation}
\langle M_V \rangle = 0.68\pm 0.14,
\end{equation}

\noindent which for a mean RR Lyr\ae\ metallicity $\rm [Fe/H] \sim
-1.5$ lies in between the two scales, although not
formally inconsistent with either.  However, Layden, Hanson, \& Hawley (1994,
in preparation) have gathered a larger sample, and  their data are
reportedly compatible with the Carney \et\ (1992) faint relation.  As
for the error bars attached to  these techniques, Clementini \et\ (1992)
quote a typical uncertainty for the surface brightness and IR flux
versions of $\pm 0.20$ mag, about the size of the discrepancy between the two
scales. The statistical parallax techniques have typical uncertainties
$\pm 0.15$ mag; however in both cases the reason for a systematic effect
that would cause an offset toward fainter magnitudes is not obvious.
Fernley (1994) has argued for an upward revision of the BW luminosities,
still leaving a discrepancy over 0.1 mag with the bright distance
scale, but moving the BW results in the right direction.

In more direct support of the brighter distance scale implied by
models with `canonical' helium abundance is the LMC RR Lyr\ae\ data of Walker
(1992). He
studied 182 variables in 7 clusters. Assuming the Cepheid distance
scale to the Cloud, $(m-M) = 18.55$ gives $\beta = 0.73,$
consistent within the errors with the theoretical tracks of higher
$Y.$ The Walker zero point appears to be supported by data from
three M31 globulars observed by HST (Faber 1995, private communication).

As well, for the main-sequence fitting data studied by Carney
\et\ (1992, their Table 1), applying their corrections to the main
sequence fitting-derived distances implies the brighter distance scale
with $\beta\sim 0.8$, rather than the relation they favour. They also
argue from the statistical parallax between the parallax subdwarf
Groombridge 1830 (HD 103095) and the globular cluster M5 in favour of
the BW zero point. However, the main sequence of M5 at the colour of
HD 103095 is rather wide owing to photometric uncertainty on the
lower main sequence.  The
distance that they quote for M5 is derived by main-sequence fitting
uncorrected for the metallicity dependent colour term they find.

To summarize, we cannot yet determine exactly how bright
the HB stars are directly from the RR Lyr\ae\ observations, but
reconciling the remaining discrepancies between the two possible
cluster distance/age scales that result is a high priority. As is also
the case with the faint blue HB stars found in $\omega$ Centauri
(Fig~\ref{fig:wcen}), resolution of the difference in favour of the
observations  may yet require some important rethinking of HB
%
%
\section*{Acknowledgements}
I would like to acknowledge a Research Associateship from the US
National Research Council, and support from NASA RTOP 188-41-51-03. I
would also like to express appreciation for helpful discussions with
Forrest Rogers, Carlos Iglesias, Allen Sweigart, Bob Rood, and Ruth
Peterson.

%
%

\beginrefer

\def\journal#1#2#3#4#5{\refer{{#1,} {#2,} {\sl #3}, {\bf #4}, #5}}

\def\jsub#1#2#3{\refer{{#1,} {#2,} {\sl #3}, submitted}}
\def\jpress#1#2#3{\refer{{#1,} {#2,} {\sl #3}, in press}}
\def\refline{\hbox to 2.truecm{\leaders\hrule height 1pt
depth -0.6pt \hfill \hskip 2pt}\nobreak}

{

\journal{Barnes, T. G., \& Hawley, S. L.}{1986}{ApJ}{307}{L9}

\journal{Bressan, A., Chiosi, C., \& Fagotto, F.}{1994}{ApJS}{94}{63}

\journal{Briley, M. M., Hesser, J. E., Bell, R. A., Bolte, M. J., \& Smith,
G. H.}{1994}{AJ}{108}{2183}

\journal{Brocato, E., Matteucci, F., Mazzitelli, I., \& Tornamb\`e,
A.}{1990}{ApJ}{349}{458}

\journal{Buonanno, R., Corsi, C. E., \& Fusi Pecci, F.}{1989}{A\&A}{216}{80}

\journal{Buonanno, R., Corsi, C. E., Gratton, R., Caloi, V., \& Castellani, V.}
{1986}{A\&AS}{66}{79}

\journal{Buzzoni, A., Fusi Pecci, F., Buonanno, R., \& Corsi, C. E.}{1983}
{A\&A}{128}{94}

\journal{Burstein, D., Bertola, F., Buson, L., Faber, S. M., \& Lauer, T. R.}
{1988}{ApJ}{324}{440}

\journal{Cacciari, C. \& Freeman, K. C.}{1983}{ApJ}{268}{185}

\journal{Castellani, M., \& Tornamb\`e, A.}{1991}{ApJ}{381}{393}

\journal{Castellani, V., \& de~Santis, R.}{1994}{ApJ}{430}{624}

\journal{Carney, B., Storm, J. \& Jones, R. V.}{1992}{ApJ}{386}{663}

\journal{Catelan, M. \& de Freitas Pacheco, J.}{1995}{A\&A}{297}{345}

\journal{Caughlan, G., \& Fowler, W.} {1988}{Atom. \& Nuc. Data Tab.} {40}{283}

\journal {Caughlan, G., Fowler, W., Harris, M. J., \& Zimmerman, B.
A.}{1985}{Atom. \& Nuc. Data Tab.}{32}{192}

\jpress {Chaboyer, B., Demarque, P., \& Sarajedini, A.}{ApJ}{1996}

\journal{Clementini, G., Cacciari, C., Fernley, J., \& Merighi, R.}{1992}
{Mem. S. A. It.}{63}{397}

\journal{Code, A. D. \& Welch, G. A.}{1979}{ApJ}{228}{95}

\journal{Cox, A.N., Hodson, S. W., \& Clancy, S. P.}{1983}{ApJ}{266}{94}

\jpress{D'Cruz, N., Dorman,  B., Rood, R. T., \& O'Connell, R. W.}{1996}{ApJ}

\jsub{Dixon, W. V., Davidsen, A. F., Dorman, B., \& Ferguson, H. C.}{1996}{AJ}

\journal{Dorman, B.}{1992}{ApJS}{80}{701}

\refer{\refline\hskip2pt, 1993 in {\it The Globular Cluster-Galaxy Connection}
eds. J. P. Brodie \& G. H. Smith (San Francisco:ASP) p. 198}

\journal{Dorman, B., Lee, Y.-W., \& VandenBerg D. A.}{1991}{ApJ}{366}{115}

\journal{Dorman, B.,  O'Connell, R. W., \& Rood, R. T.}{1995}{ApJ}{442}{105}

\journal{Dorman, B. \& Rood, R. T.}{1993}{ApJ}{409}{387}

\journal{Dorman, B., Rood, R. T., \& O'Connell, R. W.}{1993}{ApJ}{419}{596}

\journal{Fernley, J.}{1993}{A\&A}{268}{591}

\journal{\refline}{1994}{A\&A}{284}{L16}

\journal{Fontaine, G., Graboske, H., \& van Horn, H. M.}{1977}{ApJS}{35}{293}

\journal{Fusi Pecci, F., Cacciari, C.,  Bellazzini, M., \& Ferraro,
F.}{1995}{AJ}{110}{1664}

\journal{Fusi Pecci, F. \& Renzini, A.}{1976}{A\&A}{46}{447}

\journal{Greenstein, J. L., \& Sargent, A. I.} {1974}{ApJS}{28}{157}

\journal{Greggio, L. \& Renzini, A.}{1990}{ApJ}{360}{35}

\refer{Heber, U. 1992 in {\it The Atmospheres of Early-Type Stars}
eds. U. Heber \& C. S. Jeffery, (Springer: Berlin) p.233}

\refer{Holzer, T. E. \& MacGregor, K. B. 1985 in {\it Mass Loss from Red
Giants}
eds. M. Morris \& B. Zuckerman (Dordrecht:Reidel) p.229}

\journal{Horch, E., Demarque, P., \& Pinsonneault, M.}{1992}{ApJ}{388}{L53}

\journal{Iben, I., Jr \& Renzini, A.}{1984}{Phys. Rep.}{105}{329}

\journal{Iben, I., Jr \& Rood, R. T.}{1970}{ApJ}{161}{587}

\journal{Iglesias, C. A., Wilson, B. G., Rogers, F. J., Goldstein, W. H.,
Bar-Shalom, A., \& Oreg, J.}{1995}{ApJ} {445}{855}

\journal{Kraft, R. P., Sneden, C., Langer, G. E., \& Shetrone, M.}
{1993}{AJ}{106}{1490}

\journal{Lee, Y.-W., Demarque, P., \& Zinn R. J.}{1990}{ApJ}{350}{155}

\journal{\refline}{1994}{ApJ}{423}{265}

\journal{MacGregor, K. B. \& Stencel, R. E.}{1992}{ApJ}{397}{644}

\jsub{McLaughlin, D. E., \& Pudritz, R.}{1996}{ApJ}

\journal{Naur, P. E. \& Osterbrock, D.}{1953}{ApJ}{117}{306}

\refer {Nemec, J. \& Matthews, J. A. 1993  {\it New Perspectives on
Stellar Pulsation and Evolution} (Cambridge:CUP)}

\journal{Peterson, R. C., Rood, R. T., \& Crocker, D. A.}{1995}{ApJ}{453}{214}

\refer{Renzini, A. 1977 in {\it Advanced Stages of Stellar Evolution}
(Geneva:Geneva Obs.), p.149}

\journal{Renzini, A., \& Fusi Pecci, F.}{1988}{ARA\&A}{26}{199}

\journal{Rogers, F. J., \& Iglesias, C. A.}{1992}{ApJS}{79}{507}

\journal{Rood, R. T.} {1973}{ApJ} {184}{815}

\refer{Saffer, R. A. \& Liebert, J. W. 1995 in {\it Proceedings of the 9th
European Workshop on White Dwarfs} eds. D. Koester \& K. Werner
(Berlin:Springer) p.221}

\journal{Sandage, A.}{1982}{ApJ}{252}{553}

\journal{\refline}{1986}{ARA\&A}{24}{421}

\journal{\refline}{1990}{ApJ}{350}{631}

\journal{Sandage, A. \& Cacciari, C.}{1990}{ApJ}{350}{645}

\journal {Schwarzschild, K.} {1906}{Gott. Nach.}{1}{41}

\refer{Schwarzschild, M. 1958 {\it The Structure and Evolution of the Stars}
(Princeton:PUP)}

\journal{Schwarzschild, M. 1958 \& H\"arm, R.}{1965}{ApJ}{142}{855}

\journal{Searle, L., \& Zinn, R. J.}{1978}{ApJ}{225}{357}

\journal{Seaton, M. J., Yan, Y., Mihalas, D., \& Pradhan, A.
K.}{1994}{MNRAS}{266}{805}

\journal{Silbermann, N. A. \& Smith, H. A.}{1995}{AJ}{109}{1119}

\journal{Smith, G. H. \& Tout, C. A.}{1992}{MNRAS}{256}{449}

\journal{Simon, N. R.}{1982}{ApJ}{260}{L87}

\refer {Sweigart, A. V. 1990 in {\it The Confrontation Between
Stellar Evolution and Pulsation}, eds.  C. Cacciari \& G. Clementini
(San Francisco:ASP) p.1}

\journal{Sweigart, A.V., \& Mengel, J. E.}{1979}{ApJ}{229}{624}

\journal{van Albada, T. S. \& Baker, N.}{1971}{ApJ}{169}{311}

\journal{VandenBerg, D. A., Bolte, M. J., \& Stetson, P.
B.}{1990}{AJ}{100}{445}

\journal{Walker, A. R.}{1992}{ApJ}{390}{L81}

\journal{Watson, A. M., \et}{1994}{ApJ}{435}{L55}

\journal{Weaver, T. A. \& Woosley, S. A.} {1993} {Phys. Rep.} {227}{65}

\journal{Whitney, J. H. \et}{1994}{AJ}{108}{1350}

\journal {Yi, S., Lee, Y.-W., \& Demarque, P.}{1993}{ApJ}{411}{L25}

\journal {Zinn, R. J.}{1980}{ApJ}{241}{602}

}
\endrefer

\noindent{\bf E. Schatzman:} What are the masses and radii of the fast rotators
(40 kms$^{-1}$) ?
\vglue 12truept

\noindent{\bf B. Dorman:} Typical masses for blue HB stars are about 0.6 \msun.
The stars in question have surface temperatures 10000 K, which corresponds to
between 1 and 2 $R_{\odot}.$ Rotation is slower than seen in many normal
pop I stars of similar spectral type.

\end{document}